\documentclass[usenatbib]{mn2e}
\usepackage{graphics}
\usepackage{txfonts}
\title[Time delays in PG1115+080: new estimates]
{Time delays in PG1115+080: new estimates}
\author [Vakulik V.G. et al.] { \parbox[t]{\textwidth}{
\vspace{-1.0cm}  
      V. G.Vakulik$^{1,2}$ \thanks {E-mail: vakulik@astron.kharkov.ua},
       V.M.Shulga$^2$, R.E.Schild$^3$, V.S.Tsvetkova$^2$, V.N.Dudinov$^
       {1,2}$,\\A.A.Minakov$^2$, S.N.Nuritdinov$^5$, B.P.Artamonov$^4$, 
       A.Ye.Kochetov$^{1,2}$, G.V.Smirnov$^1$, A.A.Sergeyev$^{1,2}$, 
       V.V.Konichek$^1$, I.Ye.Sinelnikov$^1$, V.V.Bruevich$^4$, 
       T.Akhunov$^5$, O.Burkhonov$^5$}
       \vspace*{6pt}\\
       $^1$Institute of Astronomy of Kharkov National University, Sumskaya
           35, 61022 Kharkov, Ukraine\\
       $^2$Institute of Radio Astronomy of Nat.Ac.Sci. of Ukraine, 
           Krasnoznamennaya 4, 61002 Kharkov, Ukraine\\
       $^3$Center for Astrophysics, 60 Garden Street, Cambridge, MA
           02138, U.S.A.\\
       $^4$Sternberg Astronomical Institute, Universitetski Ave. 13, Moscow, 
            Russia\\ 
       $^5$Astronomical Institute of Ac.Sci.of Uzbekistan, Astronomicheskaya 33,
           Tashkent, Uzbekistan\\
        \vspace*{-0.5cm}}           
            
\begin{document}      
\date{Accepted ...
      Received ...;
      in original form ...}

\pagerange{\pageref{firstpage}--\pageref{lastpage}}
\pubyear{...}

\maketitle

\label{firstpage}

\begin{abstract}
We report new estimates of the time delays in the quadruple gravitationally 
lensed quasar PG1115+080, obtained from the monitoring data in filter $R$ 
with the 1.5-m telescope at the Maidanak Mountain (Uzbekistan, Central Asia) 
in 2004-2006. The time delays are 16.4 days between images C and B, and 12 
days between C and A1+A2, with image C being leading for both pairs. The 
only known estimates of the time delays in PG1115 are those based on 
observations by Schechter et al. (1997) -- 23.7 and 9.4 days between 
images C and B, C and A1+A2, respectively, as calculated by Schechter et al., 
and 25 and 13.3 days as revised by Barkana (1997) for the same image 
components with 
the use of another method. The new values of time delays in PG 1115+080
may be expected to provide larger estimates of the Hubble constant thus 
decreasing a diversity between the $H_0$ estimates taken from gravitationally 
lensed quasars and with other methods.

\end{abstract}


\section{Introduction}

As was first suggested by Refsdal (1964), gravitationally lensed quasars can 
potentially provide an estimate of the Hubble constant $H_0$ independent 
of any intermediate distance ladder. This can be made from measurements of 
the time delays between the quasar intrinsic brightness variations seen in 
different quasar images. The value of $H_0$ can be obtained then (within 
the adopted cosmological model) from the observed geometry of the system, 
with the known lens and source redshifts, and with the use of physically 
validated model of mass distribution in the lensing galaxy. Since a 
phenomenon of gravitational lensing is controlled by the surface density 
of the total matter (dark plus luminous), it provides a unique possibility 
both to determine the value of $H_0$ and to probe the dark matter abundance 
in lensing galaxies and along the light paths in the medium between the 
quasar and observer. 

By now the time delays have been measured in about 20 gravitationally lensed
quasars resulting in the values of $H_0$, different for different quasars, 
while remaining noticeably less than the most recent estimate of $H_0$ 
obtained in the HST Hubble Constant Key Project with the use of Cepheids - 
$H_0=72\pm8$ km s$^{-1}$ Mpc$^{-1}$ (Freedman et al. 2001). This discrepancy 
is large enough and, if the Hubble Constant is really a universal constant, 
needs to be explained. A detailed analysis of the problem of divergent $H_0$ 
estimates inherent in the time delay method and the ways to solve it can be 
found, e.g., in Keeton \& Kochanek (1997); Kochanek (2002); Kochanek \& 
Schechter (2004); Schechter (2005). 

The quadruply lensed quasars, and PG1115+080 in particular, are known to 
better suit for determining the $H_0$ value as compared to the two-image 
lenses since they provide more observational constraints to fit the lens 
model. The PG1115 source quasar with a redshift of $z_S=1.722$ is lensed 
by a galaxy with $z_G=0.31$ (Henry \& Heasley 1986; Christian et al. 1987; 
Tonry 1998), which forms four quasar images, with an image pair A1 and A2 
bracketing the critical curve very close to each other. It is the second 
gravitationally lensed quasar discovered over a quarter of century ago, 
at first as a tripple quasar (Weymann et al. 1980). Hege et al. (1980) 
were the first to resolve the brightest image component into two images 
separated by 0.48 arcsec. Further observations (Young et al. 1981; 
Vanderriest et al. 1986; Christian et al. 1987; Kristian et al. 1993) 
have provided positions of quasar images and information about the lensing 
object, which  allowed to build a model of the system (e.g., Keeton, 
Kochanek \& Seljak 1997). In particular, Keeton \& Kochanek (1997) have 
shown that the observed quasar image positions and fluxes and the galaxy 
position can be fit well by an ellipsoidal galaxy with an external shear 
rather than by a just ellipsoidal galaxy or a circular galaxy with an 
external shear. They noted that a group of nearby galaxies detected by 
Young et al. (1981) could provide the needed external shear.

The problem of determining the Hubble constant from the time delay lenses 
is known to suffer from the so-called central concentration degeneracy, 
which means that, given the measured time delay values, the estimates of 
the Hubble constant turn out to be strongly model-dependent. In particular, 
models with more centrally concentrated mass distribution (lower dark 
matter content) provide higher values of $H_0$, more consistent with the 
results of the local $H_0$ measurements than those with lower mass 
concentration towards the centre (more dark matter). 

\begin{figure*}
\resizebox{\hsize}{!}{\includegraphics{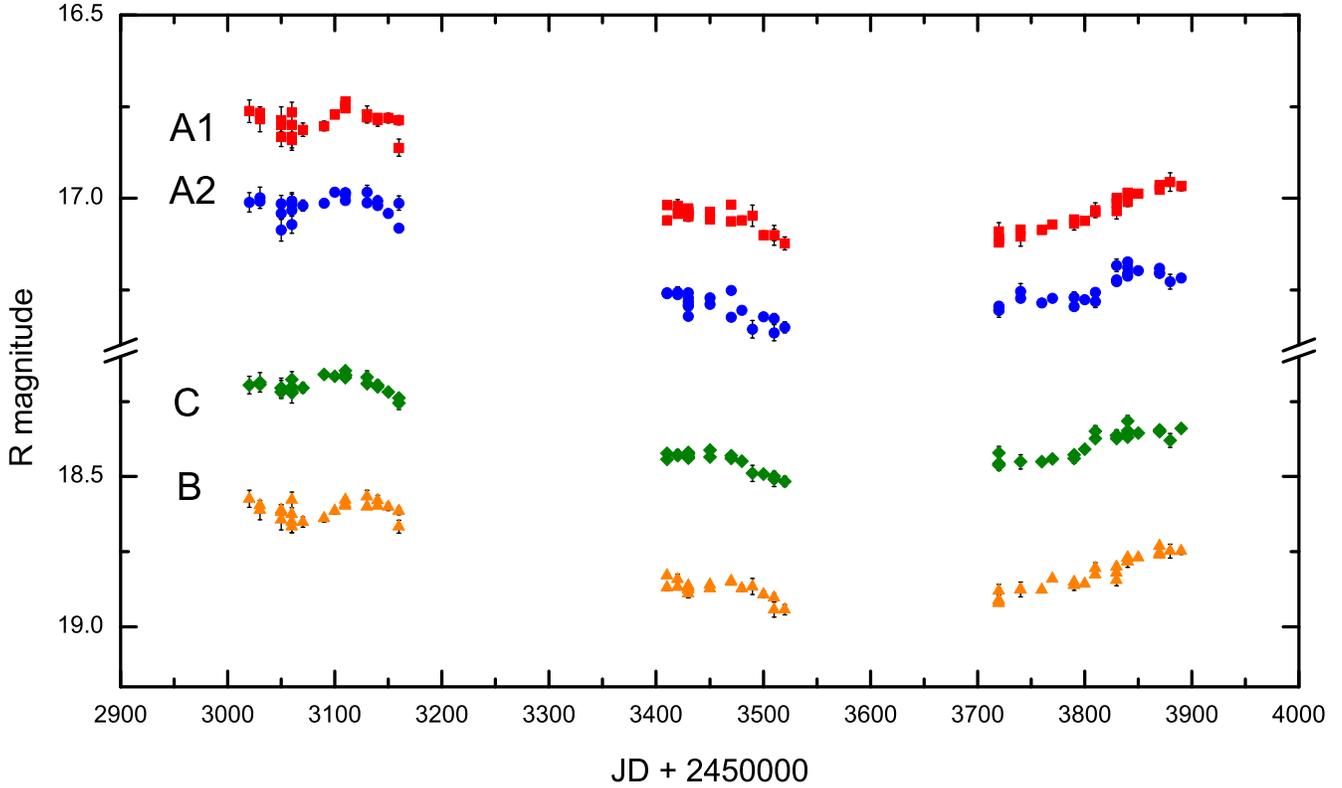}}
 \caption{The light curves of PG 1115+080 A1, A2, B, C from observations in
 filter $R$ with the 1.5-m telescope of the Maidanak Observatory in 2004, 
 2005 and 2006.}
\end{figure*}

The time delays in PG 1115+080 were determined for the first time by Schechter 
et al. (1997) to be $23.7\pm 3.4$ days between B and C, and $9.4\pm 3.4$ days 
between A1+A2 and C (image C is leading). Barkana (1997) re-analyzed their 
data using another algorithm and reported $25^{+3.3}_{-3.8}$ days for the time 
delay between B and C. There are also estimates of the time delays between 
images A1 and A2 made from the {\it Chandra} and {\it XMM-Newton} X-ray 
Observatories data (Dai X, et al. 2001, Chartas et al. 2004). As is predicted 
by all lens models, it does not exceed a small fraction of the day, and equals 
$0.16\pm0.02$ days (Chandra data) or $0.149\pm0.006$ days (XMM-Newton data). 
Determination of the time delays has generated a flow of models for the system, 
(Schechter et al. 1997; Keeton \& Kochanek 1997; Courbin et al. 1997; Impey et 
al. 1998; Saha \& Wiiliams 2001; Kochanek, Keeton \& McLeod 2001; Zhao \& Pronk 
2001; Chiba 2002; Treu \& Koopmans 2002; Yoo et al. 2005, 2006; Pooley et al. 
2006; Miranda \& Jetzer 2007), all illustrating how strongly the estimated 
value of $H_0$ depends on the adopted mass profiles of the lens galaxy for the 
given values of time delays.  

\section{Observations} 
Observations of PG 1115+080 have been carried out at the 1.5-meter telescope 
of the high-altitude Maidanak Observatory (Central Asia, Uzbekistan) with a 
scientific BroCam CCD camera. A SITe ST005A 2030 x 800 chip provided an image 
scale of 0.26\arcsec/pix at the f/8 focal plane. The CCD images were usually 
taken in series consisting of 2 to 10 frames for the $R$ filter and of 2 to 6 
frames for $V$ and $I$. To provide higher photometric accuracy, we averaged 
the values of magnitudes estimated from individual frames. The seeing varied 
from 0.75\arcsec \, to 1.3\arcsec \, (the FWHM of images of the reference 
stars B and C according to designation by Vanderriest et al. 1986). The 
analysis of photometry shows no significant dependence of the photometry 
errors on seeing, excepting the FWHM noticeably exceeding 1.3\arcsec. 
Occasional frames with such values of the FWHM were excluded from processing. 

The algorithm of photometric image processing is similar to that applied to 
photometry of Q2237+0305 and described in great details in (Vakulik et al.
2004). The light curves of PG 1115+080A1,A2,B,C in filter $R$ for three 
observational seasons in 2004-2006 are shown in Fig.1. The data sets consist 
of 23 nights in 2004, 27 in 2005, and 24 in 2006.  The data demonstrate 
noticeable variations of the quasar luminosity, with the total amplitude 
reaching almost 0.4 mag during a year (between April 2004 and May 2005). 
The time delays between the light curves of the C and B, C and A1 (or A2) 
images can be easily seen from a simple visual inspection of the $R$ light 
curves. Thus, our photometry in filter $R$  has immediately demonstrated 
applicability of the obtained light curves to determine the time delays. 
As compared to the data used by Schechter et al. (1997) and Barkana (1997), 
we were lucky to detect the quasar brightness variations with an amplitude 
of almost a factor of three larger, and with rather well-sampled data points 
within every season of observations. In addition, the accuracy of our 
photometry has made it possible to confidently detect flux variations with 
an amplitude as small as 0.06 mag, that can be seen in the data of 2004.
 
Also, recently published by Morgan et al. (2008) observations of PG1115+080 
in filter $R$ during the same time periods in 2004-2006 should be mentioned
here. We have made use of their photometry presented in their table 3 to 
compare with our light curves. Variations of the quasar brightness which 
allowed us to determine the time delays are seen in their A1+A2 light curve 
quite well, but become undetectable in  the B and C light curves because 
of a much larger scatter of the data points.

\section{Calculation of the time delays}

 The ideology of methods to determine the time delays between two image 
 components from their light curves is simple enough and obvious. 
 A common feature of all known methods of time delay measurements is the 
 use, in one way or another, of the cross-correlation maximum or mutual 
 dispersion minimum criteria, while the methods may differ in the algorithms 
 of the initial data interpolation. Analysis of the light curves of quasar 
 images in pairs can be also applied when a lens consists of more than two 
 images. But however, another approach seems to be more promising in this 
 case. The model source light curve can be determined from a joint analysis  
 of light curves of all image components. The individual time delays of the 
 components are then determined with respect to this model source light 
 curve jointly from a corresponding system of equations. In some cases, 
 this approach allows systematic variations in the light curves to be 
 revealed and taken into account, such as those caused by, e.g., 
 microlensing. A similar approach described earlier by Press et al. (1992) 
 and Rybicki \& Press (1992) and used  later by Barkana (1997) was also 
 applied to determine the time delays in Q2237+0305 (Vakulik et al. 2006). 

 As was noted above, a necessity to properly interpolate the unevenly sampled 
 data points in the light curves under consideration is one of the main 
 technical problems in determining the time delays. A variety of interpolating 
 functions and algorithms is used, such as polynomials of various power, 
 Legendre polynomials, etc. In some cases the low-frequency splines provide 
 good results. But unfortunately, all these interpolation procedures do not 
 contain any physical meaning. Meanwhile, the Fourier spectra of quasar 
 variability are known to be rapidly decreasing functions, (de Vries et al. 
 2005), and this is naturally explained by the finite physical sizes of 
 quasars. In particular, the quasar size is known to play a role of smoothing 
 factor in microlensing light curves, as is clearly seen from simulations. 
 Therefore, we tried to find such an algorithm to represent the source quasar 
 light curve, which would take into account the expected frequency 
 characteristics of the quasar variability and allow a relevant smoothing of 
 the observational data. To do this, it was natural to address an expression 
 that is well known in optics, radio engineering and theory of information as 
 the sampling theorem. According to this theorem, a signal with a bounded 
 spectrum can be represented accurately enough by a function:
 \begin{equation}
 f(t)=\sum_{k=-\infty}^\infty a(k\Delta t)sinc[\pi(t-k\Delta t)/\Delta t],
 \end{equation}
 Here, the function $sinc(x)$ is specified as $sinc(x)=sin(x)/x$, and $a(k
 \Delta t)$ are samples of $f(t)$ taken at a mesh with a step $\Delta t$, 
 which is determined by a boundary frequency $\Omega_{bnd}$ of the function 
 $f(t)$ spectrum: $\Delta t=1/(2\Omega_{bnd})$. Since, as was noted above, 
 Fourier  spectra of quasar variability are rapidly decreasing functions, 
 we may apply the  sampling theorem to reproduce the quasar light curve. 
 We used this approach earlier to analyze  statistics of microlensing 
 brightness variations in Q2237+0305 (Minakov  et al. 2008).
  
The same algorithm was used in this work to represent the model for the 
source light curve. Since the time delay between the light curves of 
images A1 and A2 is very short (Dai et al. 2001, Chartas et al. 2004), 
their fluxes were summed to form a single curve, which we will call the 
A light curve. Thus, we may write the following functional for three 
light curves:
\begin{equation}
 \Phi(\Delta t, \tau_0, \tau_1, \tau_2)={1\over {3N}}\sum_{j=0}^2
 \sum_{i=0}^N{{[m_j(t_i)+dm_j-f(t_i,\Delta t, \tau_j)]^2}\over{
 \sigma^2_j(t_i)}},
\end{equation}
where $m_j(t_i)$ are the data points in the light curve of the $j$-th image at
the time moments $t_i$, $dm_j$ and $\tau_j$ are the shifts of corresponding 
light curves in stellar magnitude and in time, $N$ is a number of points in the 
light curves, $\sigma^2_j(t_i)$ are the photometry errors, and finally, $f(t_i,
\Delta t, \tau_j)$ is an approximating function (1).
 
We adopted $dm_0=0$ and $\tau_0=0$ in our calculations, that is, we fitted
the light curves of the two other images to the A light curve, and thus, $dm_1$
and $dm_2$ are the magnitude differences A-B and A-C, respectively.
At given values of $\tau_1$ and $\tau_2$, we minimize $\Phi(\Delta t, \tau_1, 
\tau_2)$ in $dm_j$ and in coefficients $a(k\Delta t)$ of the sampling function.
The values of minimum of $\Phi(\Delta t, \tau_1, \tau_2)$  were being looked 
for at a rectangular mesh $\tau_1, \tau_2$ with a step of 0.5 days in 
preliminary calculations, and of 0.2 days at a final stage. The values of  
$\tau_1, \tau_2$ corresponding to the minimal value of $\Phi(\Delta t, \tau_1, 
\tau_2)$, were adopted as the estimates of the time delays $\tau_{BA}$ and 
$\tau_{AC}$. The time delay $\tau_{BC}$ is not an independent quantity in our 
method, and can be determined as a linear combination $\tau_{BC}=\tau_{BA}+
\tau_{AC}$. 

Having calculated the time delays, we analyzed deviations $\delta_j$ of light 
curves of each image from the approximating function:
\begin{equation}
\delta_j(t_i)=m_j(t_i)+dm_j-f(t_i,\Delta t,\tau_j).
\end{equation}

We revealed a linear trend in deviations of the A light curve, and small 
parabolic trends for images B and C. We interpreted these trends as the 
effects of microlensing, which are expected to be rather slow and weak. 
We then subtracted a half of these trends from the initial light curves 
of the components, obtained new estimates of the time delays, and analyzed 
the residual trends again. This procedure continued iteratively until 
the trends became insignificant. The final estimates of $\tau_1$ and 
$\tau_2$ are obtained with the linear trend 0.0105 mag/year subtracted 
from the image A light curve, and with the parabolic trend with the 
amplitude of $\pm 0.01$ mag subtracted from the B and C light curves. 
Thus, our estimates of the time delays are $\tau_{BA}=4.4^{+3.2}_{-2.4}$, 
$\tau_{AC}=12.0^{+2.4}_{-2.0}$, and $\tau_{BC}=16.4^{+3.4}_{-2.4}$ days, 
 with the relationship $\tau_{AC}/\tau_{BA}$, more consistent with that 
determined by Barkana (1997) than by Schechter et al. (1997).
 
 To evaluate the errors of estimating the time delays and reliability of our 
 estimates, we fulfilled a numerical simulation. We selected a function 
 $f(t_i, \Delta t, \tau_j)$ used to approximate our light curves with 
 $\Delta t=0.08$ yr, as a model source light curve. The model light curves 
 of the components were obtained by shifting $f(t_i, \Delta t, \tau_j)$ by 
 the proper time delays $\tau_1, \tau_2$ and magnitude differences, and by 
 adding some random quantities to imitate the photometry errors. The 
 estimates of these errors were obtained from the analysis of deviations of 
 the observed data points from the approximating function resulting in the 
 values of standard deviations 0.008, 0.016 and 0.011 mag for images A, B 
 and C, respectively. Since the method we used might be susceptible to the 
 mutual locations of data points in the actual light curves, we selected  
 the model samples exactly at the same time moments as in the actual light 
 curves. We simulated two cases: $\tau_{BA}=14.3$ and $\tau_{AC}=9.4$ 
 days  
 as determined by Schechter et al. (1997), and $\tau_{BA}=4.4$ and $\tau_
 {AC}=12$ days (our result). We simulated 2000 random light curves 
 synthesized as described above, and calculated the resulting time delays 
 using the procedure, which was exactly the same as in the analysis of the 
 actual light curves. Admitting that we could be mistaken in selection of 
 the Nyquist interval, we fulfilled the model calculations with a more 
 low-frequency function of the source brightness variations ($\Delta t=0.12$ 
 years). No systematic biases larger than 0.3 days in the estimates of 
 simulated time delays $\tau_1$ and $\tau_2$ were revealed in both cases. 
 The results of simulations were used to estimate the 95-percent confidence 
 intervals. 
\begin{figure*}
\resizebox{\hsize}{!}{\includegraphics{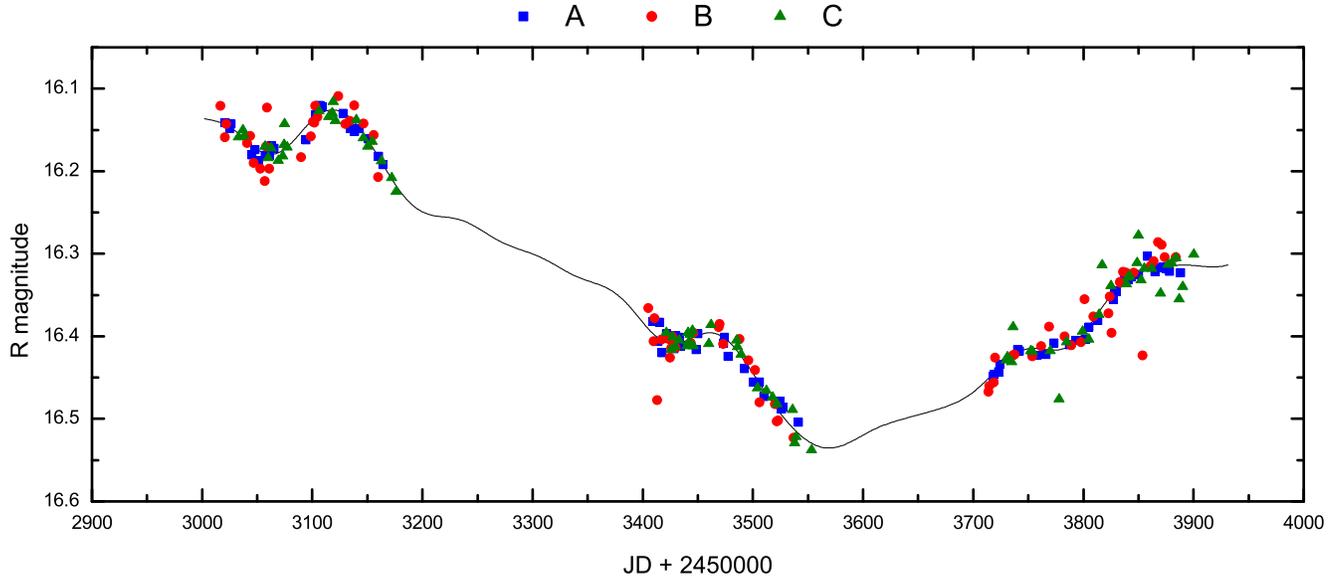}}
 \caption{The data points of the A, B and C light curves superposed with 
 each other  after shifting by the corresponding magnitude differences 
 and time delays  $\tau_{BA}=4.4^{+3.2}_{-2.5}$,  $\tau_{AC}=12.0^{+2.5}_
 {-2.0}$ and  $\tau_{BC}=16.4^{+3.5}_{-2.5}$ obtained in this work. 
 The parameter $\Delta t$  of the approximating function (shown in solid 
 curve) is 0.12 years.}
 \end{figure*}

 It is interesting to note that using only the data of 2004, where a 
 small-amplitude turn-over in the light curves is detected, we obtained
 $\tau_{BA}=5.0$, $\tau_{AC}=9.4$, and $\tau_{BC}=14.4$ days, consistent 
 with the estimates obtained from the whole data set. But  however, 
 simulation of errors for only this time interval demonstrates noticeably 
 larger uncertainties, as compared to those calculated from the whole 
 light curve. 
   
 In Fig. 2, the light curves of images A, B and C shifted by the 
 corresponding time delays and reduced to image A in magnitude are shown 
 for the approximating function parameter $\Delta t=0.12$ years.
 As is seen, the data points for all the three images are very well 
 consistent with each other and with the approximating curve, (a behavior 
 of the approximating function within the gaps between three seasons of 
 observations should be ignored).
 
So we obtained the time delay values, differing noticeably from those
by Schechter et al. (1997) and Barkana (1997), which are used in a variety 
of models of many authors to derive the Hubble constant value. Calculation 
of new lens model or recalculation of some most popular ones to derive 
the new estimate of the Hubble constant would be well beyond the scope of 
the present short communication. We would just note that the new values of 
time delays reported in this work must result in higher values of $H_0$ 
as compared to those obtained with the previous time delay values for 
PG1115+080, thus decreasing a well-known diversity between the time-delay
method of determining the Hubble constant and other methods.

\section*{Acknowledgments} 
The authors from Ukraine and Uzbekistan are grateful to the Science and 
Technology Center in Ukraine (STCU grant U127) which had made possible 
observations at the Maidanak Observatory and the further working on the 
data. The Ukrainian authors thank also the National Comprehensive 
Program "Cosmomicrophysics", and the authors from Uzbekistan are thankful
to the DFG grant 436 Usb 113/5/0-1. This work has been also supported 
by the Russian Foundation for Basic Research (RFBR) grant No.09-02-00244 
and No.06-02-16857.
 

\end{document}